# Room-temperature lasing action in GaN quantum wells in the infrared 1.5 μm region

V. X. Ho[1], T. M. Al tahtamouni[2], H. X. Jiang[3], J. Y. Lin[3], J. M. Zavada[4], N. Q. Vinh[*,1]

[1]Department of Physics & Center for Soft Matter and Biological Physics, Virginia Tech, Blacksburg, Virginia 24061, USA

[2]Materials Science & Technology Program, College of Arts & Sciences, Qatar University, Doha 2713, Qatar

[3]Department of Electrical and Computer Engineering, Texas Tech University, Lubbock, Texas 79409, USA

[4]Department of Electrical and Computer Engineering, New York University, Brooklyn, New York 11201, USA

KEYWORDS: Silicon, GaN, Quantum Wells, Lasing, Infrared Laser, Rare Earth

**ABSTRACT:**

Large-scale optoelectronics integration is strongly limited by the lack of efficient light sources, which could be integrated with the silicon complementary metal-oxide-semiconductor (CMOS) technology. Persistent efforts continue to achieve efficient light emission from silicon in the extending the silicon technology into fully integrated optoelectronic circuits. Here, we report the realization of room-temperature stimulated emission in the technologically crucial 1.5 μm wavelength range from Er-doped GaN multiple-quantum wells on silicon and sapphire. Employing the well-acknowledged variable stripe technique, we have demonstrated an optical gain up to 170 cm$^{-1}$ in the multiple-quantum well structures. The observation of the stimulated emission is accompanied by the characteristic threshold behavior of emission intensity as a function of pump fluence, spectral linewidth narrowing and excitation length. The demonstration of room-temperature lasing at the minimum loss window of optical fibers and in the eye-safe wavelength region of 1.5 μm are highly sought-after for use in many applications including defense, industrial processing, communication, medicine, spectroscopy and imaging. As the synthesis of Er-doped GaN epitaxial layers on silicon and sapphire has been successfully demonstrated, the results laid the foundation for achieving hybrid GaN-Si lasers providing a new pathway towards full photonic integration for silicon optoelectronics.

Driven by the strong need for cheap and integrable Si-based optoelectronic devices for a wide range of applications, continuing endeavors have been made to develop structures for light emission, modulation, and detection in this material system. Recent breakthroughs including the demonstration of a high-speed optical modulator in Si,[1-2] photodetectors[3], and waveguides[4] have brought the concept of transition from electrical to optical interconnects closer to realization. However, the base for silicon photonics, namely, a group IV laser source, still has to be developed. Due to the relatively small and indirect band gap, silicon is a poor light emitter. Nevertheless, lasing devices based on Si have been demonstrated including Si-based impurity lasers,[5] a Raman laser,[6] Si nanocrystals,[7] nanopatterned crystalline Si,[8] GeSn alloy on Si,[9] Ge dots in Si,[10-11] and InGaAs/GaAs nanolasers grown on Si.[12] However, these prototype devices essentially lack the advantages associated with the silicon system by requiring an external pump laser source or function only at low temperatures. While room-temperature luminescence has been realized,[7, 13] population inversion and optical gain have been under discussion and fundamental problems remain.

The incorporation of rare earth elements into semiconductor hosts gives rise to sharp, atomic-like and temperature independent emission lines under either optical or electrical excitation.[14-18] Er ions with intra-4$f$ shell transitions from its first excited state ($^4I_{13/2}$) to the ground state ($^4I_{15/2}$) produce 1.5 $\mu$m emission which falls within the minimum loss window of optical fibers for optical communications and in the eye-safe wavelength region. Lasers operating around 1.5 $\mu$m are highly sought-after for use in defense, industrial processing, medicine, spectroscopy, imaging and various other applications where the laser beam is expected to travel long distances in free space. With a tremendous effort, 1.5 $\mu$m emission from Er doped narrow-bandgap semiconductors including Si and GaAs has a low efficiency at room-temperature due to the strong thermal quenching effect.[15, 19]

**RESULTS AND DISCUSSION**

Recently, we have successfully synthesized GaN:Er epilayers on Si (001)[20-21] and c-plane sapphire[22] substrates by metal organic chemical vapor deposition (MOCVD) with excellent material qualities. The above host bandgap[20-23] and electroluminescence[24] excitation of Er optical centers produced predominant light emission at 1.5 $\mu$m range. In order to overcome the challenges of growth of III-nitrides on Si (100) substrate due to the different crystalline structures between GaN and Si, we have employed selective area growth and epitaxial lateral overgrowth techniques to prepare GaN/AlN/Si (100) templates.[20] The X-ray diffraction and photoluminescence (PL) measurements indicated that GaN:Er epilayers grown on Si and sapphire have high crystallinity, without second phase formation,[20-22] and exhibit a strong room-temperature emission at 1.5 μm with a low degree of thermal quenching.[22-23] In this work, a set of 200-period Er-doped GaN/AlN multiple quantum wells (MQWs:Er) produced a significant improvement of the quantum efficiency of the 1.5 μm emission via carrier quantum confinement and strain engineering[25] (see schematic in Fig. 1a). The growth process was started with AlN buffer and template layers and then followed by the growth of the MQWs:Er. The structure consists of alternating layers of Er doped GaN quantum wells and undoped AlN barriers (see Methods and Supporting Information). A detail description of the growth process and epilayer structure has been reported previously.[25]

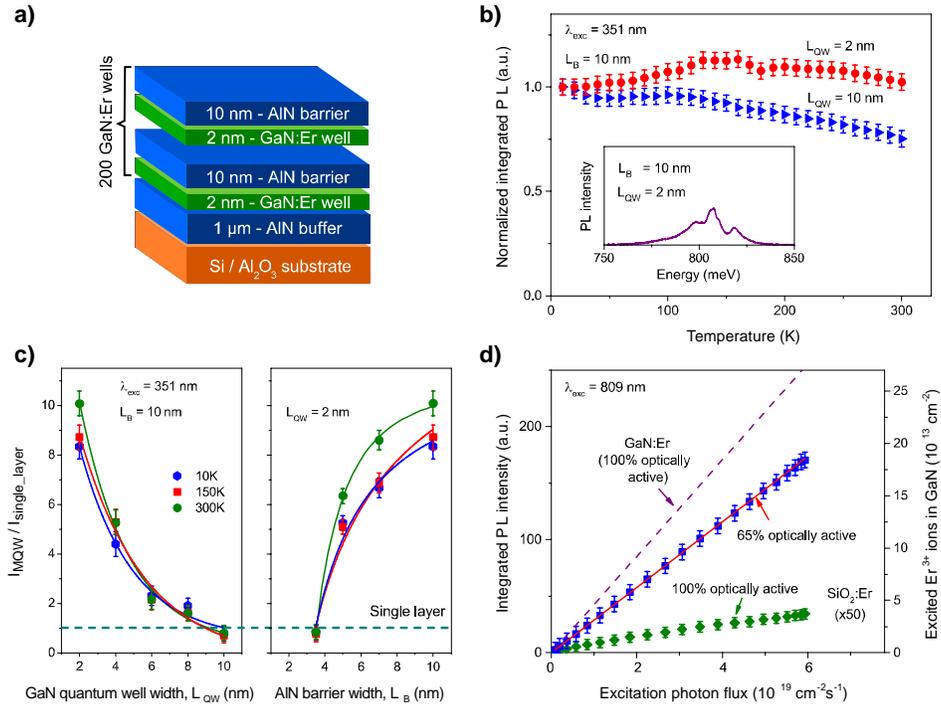

**Figure 1: Schematic of Er doped GaN/AlN MQWs and PL intensity data at 1.5 μm. a,** A 200-period MQWs:Er sample with the GaN quantum well width, $L_{QW}$, of 2 nm and the AlN barrier thickness, $L_B$, of 10 nm. **b,** The temperature dependence of PL intensity for two samples with the same the barrier thickness, $L_B$ = 10 nm, but two different quantum well widths, $L_{QW}$ = 2 nm and 10 nm, under the band-to-band excitation, $\lambda_{exc}$ = 351 nm. Inset: a typical PL spectrum from MQWs:Er materials and a single GaN:Er epilayer collected from the surface showing the spontaneous emission process. **c,** Integrated PL intensity ratio of MQWs:Er samples to a single GaN:Er epilayer at 1.5 μm as a function of the quantum well width, $L_{QW}$ (left) and the barrier thickness, $L_B$ (right), measured at different temperatures from 10 K to room-temperature under the over bandgap excitation. The maximum PL intensity was obtained with a MQWs:Er sample having $L_{QW}$ = 2 nm and $L_B$ = 10 nm. **d,** Comparison of the integrated PL intensity of $Er^{3+}$ ions in MQWs with the reference $SiO_2$:Er sample under the resonant ($^4I_{15/2} \rightarrow {}^4I_{9/2}$) excitation, $\lambda_{exc}$ = 809 nm. Measurements indicated that ~ 65% of $Er^{3+}$ ions are optically active centers in the MQWs:Er sample with $L_{QW}$ = 2 nm and $L_B$ = 10 nm.

Typical room-temperature PL spectra from the MQWs:Er samples and a single GaN:Er epilayer show a broad spectral feature[25] around 1.5 µm under the band-to-band excitation using an Ar laser at 351 nm (Fig.1 b, inset). The PL spectrum was collected from the surface of the sample, which is the spontaneous emission of light from Er optical centers in MQWs. As previously reported, the full-width at half-maximum (FWHM) of the 1.5 µm emission at room temperature is 50 nm for a single GaN:Er epilayer[22-23] and 60 nm for MQWs:Er materials.[25] The broadening of the emission in MQWs is due to the fluctuation of the GaN quantum well width and the local atomic structures around Er optical centers. This is especially a factor for Er optical centers located close to the quantum well/barrier interfaces.

The emission intensity from the MQWs:Er structure is an order of magnitude higher than that of a single GaN:Er epilayer with the same Er doping active layer thickness and concentration under the over bandgap excitation, $\lambda_{exc}$ = 351 nm. The strong quantum confinement effects on the Er emission have been studied for different values of quantum wells and barriers. In Fig. 1c, the integrated PL intensity ratio of MQWs:Er samples, $I_{MQW}$, to a single GaN:Er epilayer, $I_{single\_layer}$, at 1.5 µm is shown as a function of the GaN quantum well width, $L_{QW}$, (left) and the AlN barrier thickness, $L_B$, (right), measured at temperatures from 10 K to room-temperature. As shown in the Fig. 1c (left) for MQWs:Er samples with $L_B$ = 10 nm, the integrated PL intensity, $I_{MQW}$, increases significantly when $L_{QW}$ was reduced from 10 nm to 2 nm. In these measurements, $I_{MQW}$ is normalized to the integrated PL intensity of the single GaN:Er epilayer, $I_{single\_layer}$, with the same total Er doping active layer thickness and concentration. When the quantum well width, $L_{QW}$, is larger than 10 nm, $I_{MQW}$ is approximately equal to $I_{single\_layer}$ (the horizontal dashed lines in Fig. 1c). A decreasing of the quantum well width provides a strong quantum confinement effect of carriers around Er ions, thus improves the quantum efficiency of the 1.5 µm emission from Er ions in GaN. When the quantum well width is smaller than the free exciton Bohr radius of 2.8 nm in GaN,[26] the built-in electrical fields due to the lattice mismatch between GaN/AlN produce further the strong quantum confinement effect for excitons, resulting in an efficient energy transfer from excitons to Er optical centers. The highest integrated PL intensity has been obtained for $L_{QW}$ = 2 nm (Fig. 1c, left). For MQWs:Er samples with $L_{QW}$ = 2 nm, the integrated PL intensity, $I_{MQW}$, is enhanced by more than one order of magnitude over that of the single GaN:Er epilayer via the variation of the AlN barrier thickness, $L_B$, (Fig. 1c, right). When the $L_B$ is larger than the exciton Bohr radius in GaN, electron wavefunctions are localized at the quantum well. A large AlN barrier thickness provides an increased probability of capturing excitons by Er optical centers, leading to a higher excitation efficiency of $Er^{3+}$ ions. We have obtained the maximum $I_{MQW}$ at 1.5 µm when the AlN $L_B$ is above 10 nm.

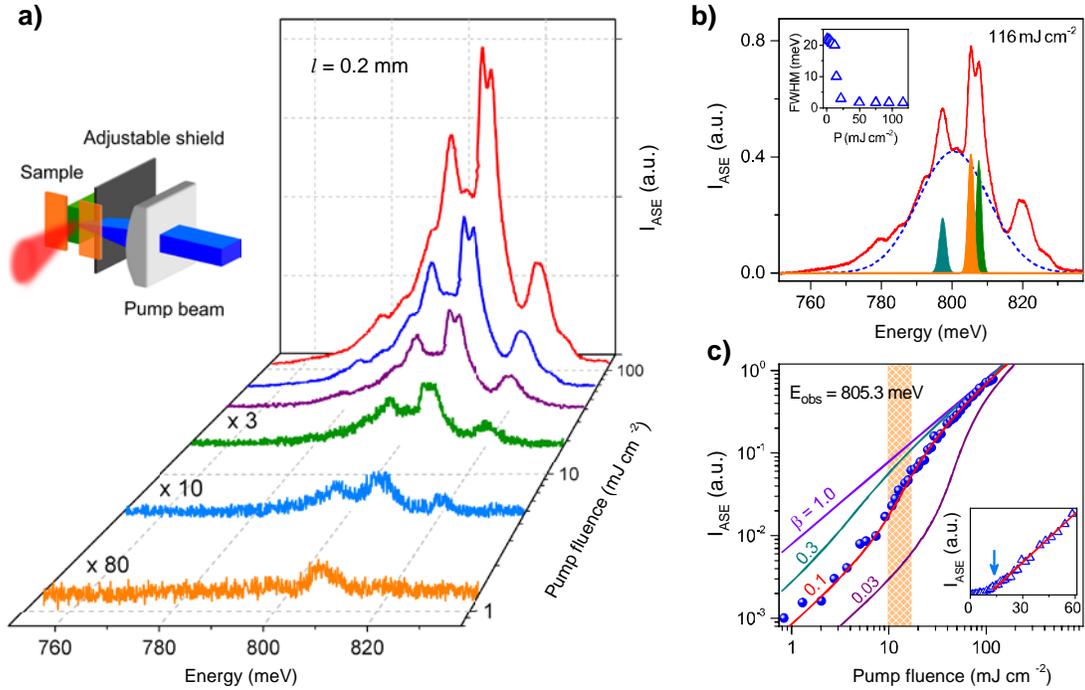

**Figure 2: Amplified spontaneous emission, at room-temperature, from the MQWs:Er sample with $L_{QW}$ = 2 nm and $L_B$ = 10 nm using an Ar laser, $\lambda_{exc}$ = 351 nm, for the band-to-band excitation. a,** Pump-fluence dependent PL spectra obtained with a 8 μm wide and 0.2 mm long pump excitation. At low excitation pump fluence, the emission is broad with the FWHM of 30 meV. When the pump fluence is high enough for the optical layer to have net gain, the spontaneously emitted photons are exponentially amplified by stimulated emission and the spectral peaks become narrower. Inset: Schematic of experimental configuration for edge-emission and variable excitation length measurements. **b,** PL spectrum at room-temperature with a pump fluence of 116 mJ cm$^{-2}$ and an excitation length of 0.2 mm. The spectrum was deconvolved into 6 Gaussian peaks. The most prominent modes show a FWHM of 1.60 ± 0.25 meV. Inset: The FWHM of the Gaussian peaks decreases with increasing the pump fluence. **c,** L-L data on a log-log scale showing the dependence of the edge-emission intensity for PL peak at 805.3 meV on the pump fluence. The excitation length was 0.2 mm and the data were fitted using the S-curve model. Inset: The edge-emission intensity showing linear behavior below threshold and superlinear increase at higher pump fluence.

In order to evaluate the thermal quenching effect of the 1.5 μm emission, we have performed the temperature dependence of the integrated PL intensity of MQWs:Er samples under the band-to-band excitation, $\lambda_{exc}$ = 351 nm. The PL experiments were carried out in a variable temperature closed-cycle optical cryostat (Janis) providing a temperature range from 10 K to 300 K. The integrated PL intensity of Er in the single GaN:Er epilayer, $I_{single\_layer}$, is reduced by about 20% from 10 K to room-temperature (Fig. S3a, Supporting Information).[23] The integrated PL intensity of Er in MQWs:Er samples, $I_{MQW}$, with quantum well width larger than the exciton Bohr radius in GaN and the AlN barrier thickness thinner than 10 nm follow the same temperature dependence behaviors as the single GaN:Er epilayer, $I_{single\_layer}$, (Fig. 1b). However, when the thicknesses of the quantum wells and barriers are close to the optimal parameters ($L_{QW}$ = 2 nm, $L_B$ = 10 nm), the temperature dependence of PL intensity is quite different. The highest value of $I_{MQW}$ occurs at 150 K and the values of $I_{MQW}$ at 10 K and room-temperature are comparable (Fig. 1b). When the temperature increases, the mobility of carriers in GaN increases significantly and reaches a maximum around 130 K.[27-28] The increasing of the carrier mobility provides higher efficiency for the energy transfer from GaN quantum wells to Er ions, resulting in the stronger PL emission intensity. The contamination of Al in MQW structures can also provide a factor that affects the thermal property of the PL intensity of Er ions. We have further conducted the temperature dependence of the PL emission intensity for different MQW structures (see Supporting Information). In this work, we compare the PL intensity from MQWs:Er samples with the optimal structure, i.e. $L_{QW}$ = 2 nm and $L_B$ = 10 nm, and a GaN:Er single layer.

The percentage of Er ions that emit photons at 1.5 μm is a crucially important parameter for potential applications of Er-doped GaN since it determines the PL intensity as well as population inversion. An estimate of the number of emitting Er optical centers can be made by comparing the PL intensity of the MQWs:Er sample with that of a $SiO_2$:Er reference sample with the same shape and under the same experimental conditions. In order to avoid the quantum confinement effect for excitons and carriers, the experiment has been performed at room-temperature under the resonant excitation $^4I_{15/2} \rightarrow {}^4I_{9/2}$ transition using the Ti:Sapphire laser at 809 nm. Note that under the band-to-band excitation, several Er centers are excited including dark, bright centers. We have observed two kinds of optical centers in the single GaN:Er epilayer including the defect-related and isolated Er optical centers.[23] The PL from defect-related Er optical centers can be observed under the band-to-band excitation and at the sample temperature below 150 K. The isolated Er optical centers can be excited by both resonant and band-to-band excitation. Thus, this is the correct way to employ the resonant excitation for this estimation. The time-integrated PL intensity of these samples are collected as a function of applied photon fluxes (Fig. 1d). By comparing the PL from Er ions in the MQWs:Er sample with the reference sample of Er-doped $SiO_2$, we have estimated that the fraction of Er ions that emits photon at 1.5 μm is approximately 65% (Supporting Information). Using the same method, we found that a fraction of ~ 68% of Er ions in the single GaN:Er layer is optically active.[29] This achievement of the high percentage of Er ions that emit photons at 1.5 μm represents a significant step in realization of GaN:Er as an optical gain medium.

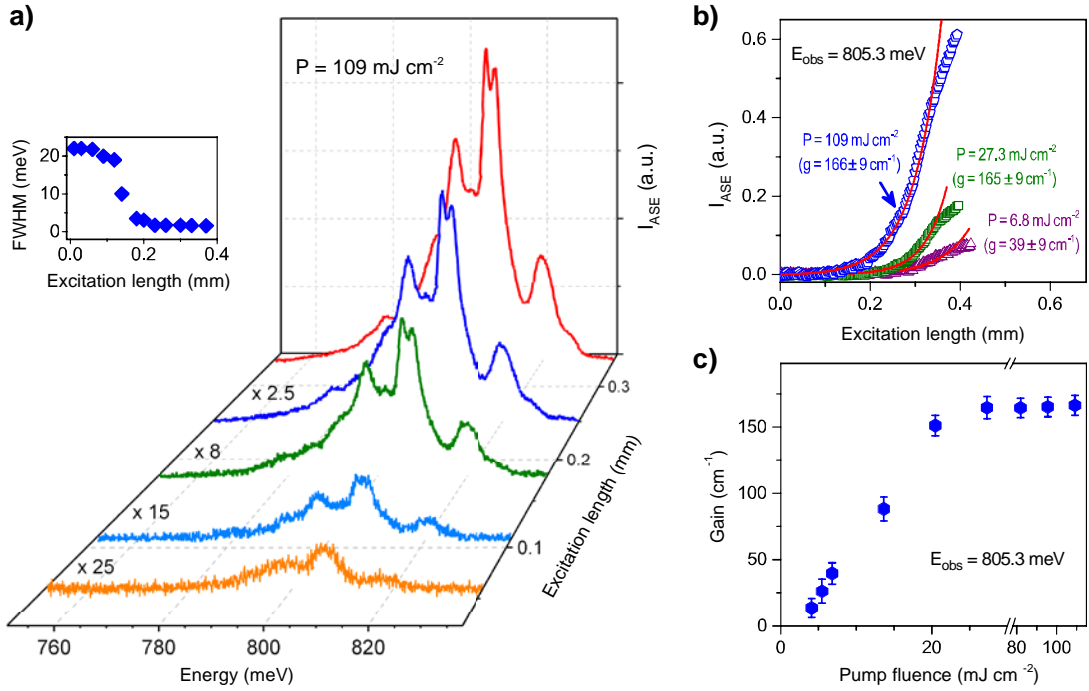

**Figure 3: Optical gain determination via the variable stripe length method. a,** Edge-emission PL spectra from the MQWs:Er sample with $L_{QW}$ = 2 nm and $L_B$ = 10 nm as a function of the excitation length at a pump fluence of 109 mJ cm$^{-2}$. The PL spectra become sharper at higher pump fluence due to the amplified spontaneous emission. The narrowing of the edge-emission spectral peaks provides evidence of lasing in the MQWs:Er sample at room-temperature. Inset: The reduction of the FWHM of the deconvoluted Gaussian peaks with increasing the excitation lengths. **b,** Dependence of the amplified spontaneous emission intensity, for the PL peak at 805.3 meV, on excitation length. The modal gain of the MQWs:Er sample was determined by using Eq. 1 to the data. **c,** The net gain of the sample at 805.3 meV as a function of the pump fluence.

The edge-emission from optically pumped MQWs:Er sample under the band-to-band excitation provides evidence of room-temperature lasing from Er optical centers in the MQW structure (Fig. 2a, inset). In order to achieve lasing from the MQWs:Er sample, both edges of the sample were polished to obtain a cavity. The edge-emission spectra from Er optical centers are different at different pump fluence near threshold (Fig. 2a). In the measurements we have employed a long excitation area with 8.0 ± 0.3 μm width and 200 ± 0.5 μm length (Fig. 2a, inset and Methods). The photon fluence of the Ar laser was varied from 0.05 to 120 mJ cm$^{-2}$. At low excitation pump fluence (P < 15 mJ cm$^{-2}$), the emission at 1.5 μm shows a broad spectrum with the FWHM of 60 nm (~30 meV) which corresponds to the spontaneous emission[25]. The broad PL spectrum is similar to those of Er ions incorporated in insulators including $SiO_2$ at room-temperature. When the pump fluence is high enough (P > 15 mJ cm$^{-2}$) for the MQWs:Er sample to obtain a net optical gain, the spontaneously emitted photons are exponentially amplified by stimulated emission as they travel through the waveguide in the active medium, leading to a superlinear increase in emission. Since the gain is maximum near the peak of the spontaneous emission spectrum, the spectrum exhibits "gain narrowing".[30-31] Consequently, an intense beam with spectral narrowing is emitted from the edge of the sample. The amplified spontaneous emission occurs at the wavelength where the spontaneous emission spectrum is strongest. The PL spectra at high excitation pump fluence indicate a number of strong and narrow PL lines. This is the signature of optical amplification of the spontaneous emission from Er optical centers around 1.5 µm in the MQW structure (Fig. 2b). We focus to analyze the strongest PL lines at 797.31, 805.30 and 807.55 meV. A deconvolution of the PL spectrum into different Gaussian components is also shown. When the pump fluencies were above 15 mJ cm$^{-2}$, the FWHM of the spectrum dropped to 1.60 ± 0.25 meV (Fig. 2b, inset). The PL intensity evolution of narrow peaks indicates a clear threshold transition from sub-threshold to linear evolution and ultimately reaches saturation at high pump-fluence (Fig. 2b).

A closer examination of this threshold behavior of the PL intensity emitted from the edge of the sample shows the typical superlinear (exponential) to linear transition, indicating the classical spontaneous-to-stimulated emission transition widely observed in semiconductor lasers.[32] Figure 2c shows the light-in-light-out (L-L) data which is the dependence of the amplified spontaneous emission on the pump fluence for the PL peak at 805.3 meV (1539.61 nm) with the excitation length of 0.2 mm. Below the threshold, the PL dependence is linear and a superlinear increase in emission intensity with pump fluence has been observed. Subsequently, stimulated emission dominates and the PL intensity evolution becomes linear (Fig. 2c, inset). Fitting the L-L data in the log-log plot to the S-curve model,[32-33] a lasing threshold of $P_{th}$ ~ 15 mJ cm$^{-2}$ has been determined (Fig. 2c), with the spontaneous emission coupling factor, $\beta$, of 0.1. The L-L curves for different $\beta$ values are also plotted for comparison, clearly showing the distinct position of the lasing threshold. Additional representative L-L plots of MQWs:Er lasing under optical pumping with different excitation lengths can be found in Supporting Information. Above the threshold for amplified spontaneous emission, most of the excitation were stimulated to emit light into waveguide modes, leading to a large fraction of the light emerged from the edge.[30, 34]

Together with the threshold behavior, we have observed the shortening of the lifetime of Er emission in the MQW structure. A typical lifetime of Er spontaneous emission in the MQW structures with the pump fluence below the threshold is around 2.5 ms (red curve, Figure S4, Supporting Information). Under a high pump fluence (above the threshold), the lifetime of Er emission shows a shortening with a dynamics < 10 µs (blue curve, Figure S4, Supporting

Information). The value is limited by the time-constant of our time-resolved photoluminescence setup. We have integrated the PL intensity for the fast and slow components. The PL spectrum for the fast component shows narrowing features as indicated in the Figure 2a at high pump fluence. The PL spectrum for the slow component is similar with the spectrum at low pump fluence. The PL intensity of the slow component can be originated from surface emission of the sample.

We have employed the well-known variable excitation length method to determine the gain coefficient from the evolution of the peak-emission intensity (Method).[35] The sample was optically excited by the Argon laser providing the band-to-band excitation in a stripe geometry. The amplified spontaneous emission signal, $I_{ASE}$, was collected as a function of the illuminated length or the excitation length, $l$, from the edge of the sample. As a result of stimulated emission, the spontaneous emitted light is amplified as it passes through the excited volume to the edge of the sample. Assuming a one-dimensional amplification model, the modal gain, $g_{mod}$, can be extracted from the amplified spontaneous emission intensity and the excitation length:[35]

$$I_{ASE}(l) = \frac{A \times I_{SPONT}}{g_{mod}}(e^{g_{mod} l} - 1) \qquad (1)$$

where $I_{SPONT}$ is the spontaneous emission intensity per unit length and A is the cross-section area of the excited volume, and the modal gain is the gain minus the loss of the material.

When the excitation length was varied, the emission spectral peaks became narrower as the excitation length was increased, and the output intensity grew exponentially. Figure 3a presents the PL spectra with an optical excitation of 109 mJ cm$^{-2}$ for different excitation lengths. The observation of an exponential increase in the intensity and a substantial decrease in spectral linewidth (Fig. 3a, inset) of light emitted as the excitation length is varied is a direct indication of the optical gain. Fig. 3b shows the output intensity at the observation peak at 805.3 meV as a function of excitation length at three different pump fluence. In each case, the data can be fitted to Eq. 1. The net gain of the waveguide was measured as a function of the pump intensity (Fig. 3c). At higher pump fluence, the output intensity increased exponentially with excitation lengths less than 0.42 mm, and then leveled off at longer excitation lengths. This behavior can be attributed to gain saturation, which occurs when the light traveling in the waveguide becomes so intense that it depletes a substantial fraction of the excited centers and reduces the gain coefficient. We limited the gain analysis to pump fluence <150 mJ cm$^{-2}$ and excitation lengths <0.42 mm.

In conclusion, our investigations provide conclusive evidence of light amplification and stimulated emission in MQWs:Er samples. In the past, this formidable goal was unsuccessfully attempted by Er doping of crystalline Si, and later of SiO$_2$ sensitized with Si nanocrystals. Here we have demonstrated the realization of this long-sought goal of Si photonics in state-of-the-art GaN MQW structures grown on-Si and GaN, demonstrating for the first time the added value provided by merging of these two most important semiconductor materials – Si for standard electronics and photovoltaics and GaN for power electronics, photonics and automotive applications.

**METHODS**

### Sample fabrication

Er doped GaN/AlN MQWs (MQWs:Er) were grown by MOCVD. The aluminum source was trimethylaluminum (TMA), the gallium source was trimethylgallium (TMGa), and the nitrogen source was ammonia ($NH_3$). Trisisopropylcyclopentadienylerbium (TRIPEr) was used as a precursor for *in-situ* Er doping. The growth started with a thin (30 nm) AlN buffer layer (buffer 1) grown at 950 °C and 30 mbar followed by a second 100 nm AlN buffer layer (buffer 2) at 1100 °C grown at 30 mbar, and a 1.0 µm AlN template grown at 1325 °C and 30 mbar. It was then followed by the growth of the MQWs:Er structure of Er doped GaN quantum wells and undoped AlN barrier layers. The growth temperature and pressure were 1000 °C and 30 mbar.

### PL intensity measurement

The PL spectra were collected with a high resolution Horiba iHR550 spectrometer equipped with a 900 grooves/mm grating blazed at 1.5 µm and detected by a high sensitivity liquid nitrogen InGaAs DSS-IGA detector. The resolution of PL spectrum is 0.05 nm. The PL experiments were carried out in a variable temperature closed-cycle optical cryostat (Janis) within the temperature range from 10 K to 300 K. We have employed resonant and non-resonant excitation to investigate optical properties from the MQWs:Er structures.[23] An Argon laser emitting light at 351 nm (3.531 eV) was employed for the non-resonant (band-to-band) excitation of MQWs:Er materials. The PL spectra under the resonant excitation from $^4I_{15/2} \rightarrow {}^4I_{9/2}$ of $Er^{3+}$ in GaN were obtained using a tunable wavelength Ti:Sapphire laser around 809 nm (1.533 eV).[23]

### Edge-emission measurements

The pump laser was magnified and then focused onto the sample's top surface using a cylindrical lens with f = 7.5 cm (Fig. 2b, inset). Only the central part of the laterally unfocused laser spot was used to excite the sample so that the pump fluence was uniform across the entire excited area. The excitation area was measured to be a long stripe of 8.0 ± 0.3 µm width exciting the entire sample length of 1000 ± 0.5 µm. We used a two-dimensional linear stage to scan our UV photodetector at the focal position of the laser to verify that within experimental conditions the laser pump fluence on the sample surface was constant and independent of the length. No influence of diffraction effects on the uniformity of the pump laser was detected. An aperture was used to insure that none of light, which has passed out of the excited volume before reaching the edge of the sample, was detected. The edge-emission was collected using a set of two lenses f = 10 cm and f = 24 cm with a diameter of 5 cm and focused onto the entrance slit of the high resolution spectrometer through a long-pass filter blocking all background light with wavelength shorter than 950 nm. All the edge-emission measurements were done at room-temperature. The photon fluence on the sample was varied from 0.05 to 500 mJ cm$^{-2}$.

**Gain measurements**

Net optical gain of the MQWs:Er sample with optimized structural parameters (i.e., GaN $L_{QW}$ = 2 nm and AlN $L_B$ =10 nm) was measured using variable excitation length method. In the same edge-emission measurement configuration, a mobile blade was mounted on an ultra-precise linear translation stage (relative accuracy of 80 nm). The edge-emission was collected and focused onto the entrance slit of the spectrometer as described. The measured gain coefficients show a fluctuation of within 5% for different distances between the sample's top surface and the mobile blade, suggesting that diffraction effects can be safely disregarded.

**AUTHOR INFORMATION**

**Correspondence Author.**

*E-mail:vinh@vt.edu

**ACKNOWLEGEMENTS**

N.Q.V. acknowledges the support from NSF (ECCS-1358564). The materials growth effort at TTU was supported by JTO/ARO (W911NF-12-1-0330).

**SUPPORTING INFORMATION**

# Room-temperature lasing action in GaN quantum wells in the infrared 1.5 μm region

V. X. Ho[1], T. M. Al tahtamouni[2], H. X. Jiang[3], J. Y. Lin[3], J. M. Zavada[4], N. Q. Vinh[*,1]

[1]Department of Physics & Center for Soft Matter and Biological Physics, Virginia Tech, Blacksburg, Virginia 24061, USA

[2]Materials Science & Technology Program, College of Arts & Sciences, Qatar University, Doha 2713, Qatar

[3]Department of Electrical and Computer Engineering, Texas Tech University, Lubbock, Texas 79409, USA

[4]Department of Electrical and Computer Engineering, New York University, Brooklyn, New York 11201, USA

*Correspondence and requests for materials should be addressed to N.Q.V.

(Email: vinh@vt.edu)

(9 pages, 4 figures)

## 1. MOCVD Growth of Er doped multiple quantum wells

For the growth of Er doped GaN/AlN multiple quantum wells (MQWs:Er), the aluminum source was trimethylaluminum (TMA), the gallium source was trimethylgallium (TMGa), and the nitrogen source was ammonia ($NH_3$). Tris(isopropylcyclopentadienyl)erbium (TRIPEr) was used for the *in-situ* Er doping. The precursors for Ga and Er were held in stainless steel bubblers at 3 °C and 60 °C, respectively, and were carried into the reactor by $H_2$ gas. The growth started with a thin (30 nm) AlN buffer layer (buffer 1) grown at 950 °C and 30 mbar followed by a second 100 nm AlN buffer layer (buffer 2) at 1100 °C grown at 30 mbar, and a 1.0 µm AlN template grown at 1325 °C and 30 mbar. It was then followed by the growth of the MQWs:Er structure of Er doped GaN quantum wells and undoped AlN barrier layers. The growth temperature and pressure were 1000 °C and 30 mbar. Figure S1 shows the temperature profile of the complete growth process.

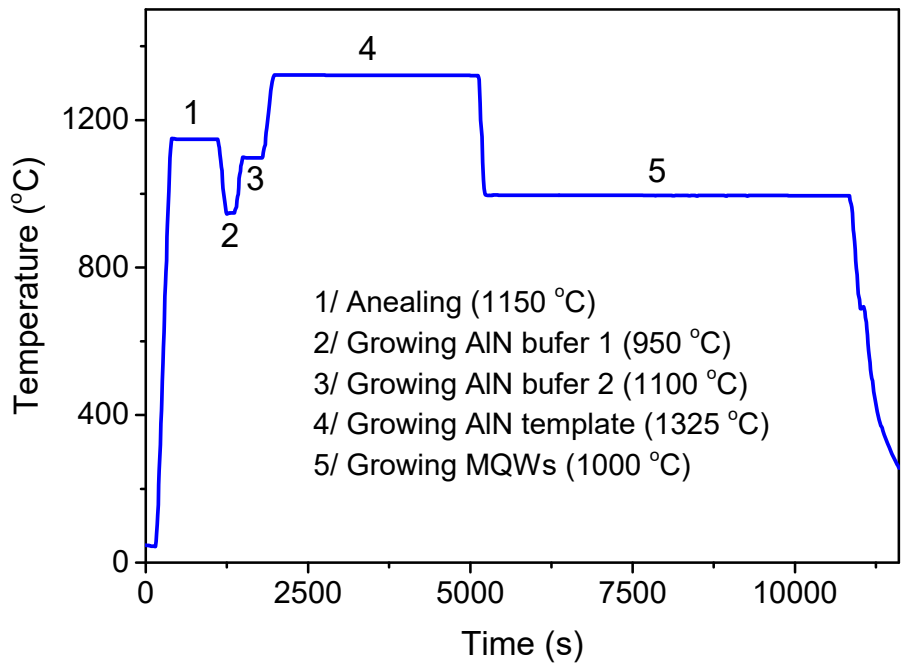

**Figure S1:** The growth temperature sequence of MQW:Er samples.

In order to evaluate the structural characteristics of the MQWs:Er, we have performed X-ray $\theta$–$2\theta$ measurements for these samples. The measurements provide information of the interfaces between the wells and barriers. The measurements have been done for a set of MQWs:Er samples with a fixed barrier thickness, $L_B$ = 10 nm, and varying quantum well width, $L_{QW}$, (0.6, 1.0, 1.5, 2.8, 4.0, 5.3, 6.6 nm).[1] Here we present the data with the same quantum well thickness of 1.5 nm, and different AlN barrier widths (1, 3, 6, 8 nm).

The peaks at $2\theta = 36.02°$ in Fig. S2 come from the (0002) plane of the underlying undoped AlN and the satellite peaks originate from the Er doped GaN/AlN MQWs. The diffraction peaks having many well-defined satellite peaks in the X-ray diffraction spectra of MQWs indicate that the interfaces between wells and barriers are abrupt.[2] As seen in Fig. S2, Er doped GaN/AlN MQWs with different barrier thicknesses exhibited several intense satellite peaks, confirming that the interfaces between quantum wells and barriers are smooth. The MQW:Er samples with barrier widths thicker than 1 nm show large numbers of intense satellite peaks, indicating that these samples possess high interfacial qualities.[1-3]

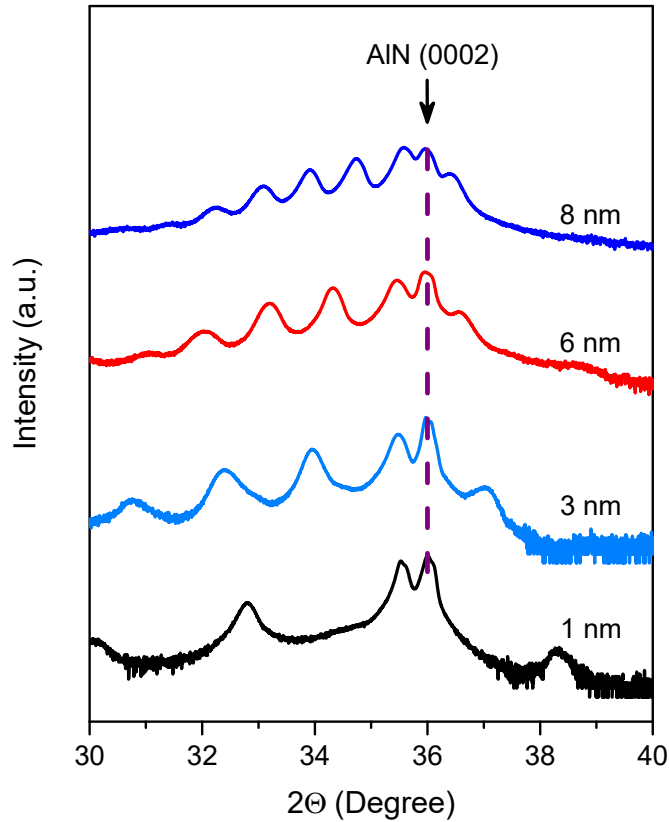

**Figure S2:** X-ray θ–2θ diffraction measurement of Er doped GaN/AlN MQWs with different AlN barrier thicknesses (1, 3, 6 and 8 nm) and fixed GaN well width of 1.5 nm.

## 2. Temperature dependence of MQWs:Er samples

In order to investigate the thermal quenching effect of the 1.5 μm emission, we have measured the integrated photoluminescence (PL) intensity of our multiple quantum wells doped with Er as a function of temperature. For MQWs with the optimal parameters with the quantum well thickness, $L_{QW}$ = 2 nm and barrier width, $L_B$ = 10 nm, we have obtained the highest PL intensity at 150 K, whereas the PL intensity at low and room-temperature are similar (Fig. S3a). Here we present the temperature dependence of the integrated PL intensity for two MQWs:Er samples with the same the GaN quantum well thickness, $L_{QW}$ = 2 nm, but different AlN barrier widths of 3.5 nm (Fig. S3a, green curve) and 10 nm (Fig. S3a, orange curve), and a GaN:Er single epilayer (Fig. S3a, red curve). In addition, we have plotted the integrated PL intensity from these samples as a function of 1000/T (Fig. S3b). As we mentioned, when the AlN barrier thickness is larger than the exciton Bohr radius in GaN, electron wavefunctions are localized at the quantum well, leading to a high excitation quantum efficiency of $Er^{3+}$ ions. Specifically, when the width of AlN barriers is larger than 10 nm, the probability of capturing excitons by Er optical centers is maximum. We have obtained an order of magnitude higher in the 1.5 μm emission intensity when compared with that from the GaN:Er single epilayer[4-5] with a comparable Er doped active layer thickness. In MQW samples with AlN barrier width of 3.5 nm, carriers are not strongly delocalized in the GaN quantum well. At elevated temperature, the mobility of carriers in GaN increases, carriers can easily escape from the GaN quantum wells, resulting in the lower intensity at high temperature. The behavior for this sample is similar with the GaN:Er single epilayer.

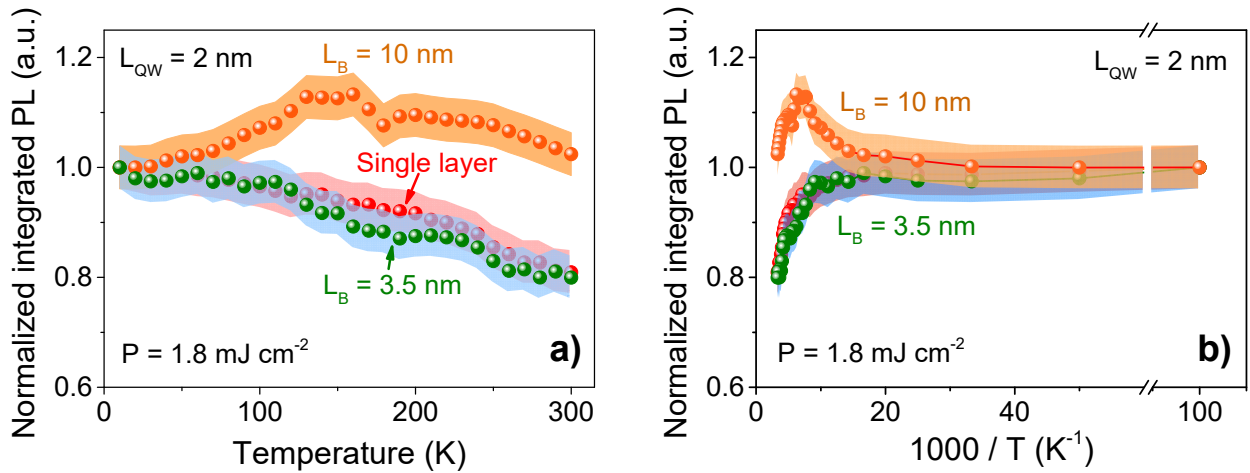

**Figure S3:** The temperature dependence of normalized integrated PL intensity for two MQWs:Er samples with the same quantum well width of $L_{QW}$ = 2 nm, but different barrier thicknesses of 3.5 nm (green color) and 10 nm (orange color), and a GaN single epilayer (red color). The measurements provide optimum parameters for the PL from MQWs:Er materials. The PL intensity reaches the maximum at 150 K for the optimum sample with the quantum well thickness of 2 nm and the barrier width of 10 nm.

## 3. Optical activity determination

The percentage of Er dopants that emit photons upon excitation is a crucially important parameter for the potential application of Er-doped semiconductors. In particular, for emission at 1.5 μm from GaN:Er materials, it determines the PL efficiency as well as population inversion. An estimate of the number of emitting Er optical centers can be made by comparing the PL intensity of the 200-period MQWs:Er grown by MOCVD technique with that of the SiO$_2$:Er reference sample under the same excitation conditions. Typically, to excite Er ions in SiO$_2$ we use resonant excitation. Thus, we should use the same excitation mechanism to compare the optical activity between two samples. In order to avoid the quantum confinement effect for excitons and carriers, the energy back-transfer and cascade transitions from higher energy levels of Er ion to the $^4I_{13/2}$, the experiment was performed at room-temperature under the resonant excitation for the $^4I_{15/2} \rightarrow {}^4I_{9/2}$ transition using the Ti:sapphire laser at 809 nm.

The time-integrated PL intensity of these samples was collected as a function of applied photon fluxes (Fig. 1d). The instantaneous PL intensities of both samples were proportional to $N_{Er}^*/\tau_{rad}$, where $N_{Er}^*$ and $\tau_{rad}$ are the density of excited Er$^{3+}$ ions and their radiative lifetimes, respectively. Since the PL signal is integrated over time, the result of the experiment will be proportional to $N_{Er}^* \times \tau/\tau_{rad}$, where $\tau$ is the effective PL lifetime. To obtain the number of photons coming out of the sample, the quantum efficiency can be determined by the refractive index of the material for a specific wavelength. We ignore in this calculation photon reabsorption, therefore, the ratio of the number of photons emitted from the two investigated samples is given by:

$$\frac{I_{GaN:Er}}{I_{SiO_2:Er}} = \frac{\eta_{GaN}}{\eta_{SiO_2}} \frac{N_{Er(GaN)}^*}{N_{Er(SiO_2)}^*} \frac{\left(\tau_{eff}^{Er(GaN)}/\tau_{rad}^{Er(GaN)}\right)}{\left(\tau_{eff}^{Er(SiO_2)}/\tau_{rad}^{Er(SiO_2)}\right)} \tag{1}$$

where η is the extraction efficiency. The ratio of the extraction efficiencies can be calculated from the refractive indexes of GaN and SiO$_2$ materials:

$$\frac{\eta_{GaN}}{\eta_{SiO_2}} = \frac{n_{air}^2/(4n_{GaN}^2)}{n_{air}^2/(4n_{SiO_2}^2)} = 0.389 \tag{2}$$

where $n_{GaN}$ and $n_{SiO2}$ are the refractive indexes of 2.318 and 1.445 at 1.5 μm for GaN and SiO$_2$, respectively.

Under resonant excitation, the decay dynamics PL from Er optical centers appear as a single exponential with a decay time constant of 3.3 ± 0.3 ms.[6] The value is similar to previous reports on Er in GaN.[7-9] The decay dynamics is effectively independent of temperature over the entire range from 10 K to room-temperature.[6] Thus, the decay time constant of 3.3 ms is the radiative lifetime, and the $\tau_{\text{eff}}^{\text{Er(GaN)}}/\tau_{\text{rad}}^{\text{Er(GaN)}} = 1$. For the SiO$_2$:Er standard sample, Er embedded in the insulating host matrix and defect free material, we obtain $\tau_{\text{eff}}^{\text{Er(SiO}_2)}/\tau_{\text{rad}}^{\text{Er(SiO}_2)} = 1$.

The well-characterized SiO$_2$:Er sample is used to relate the measured PL intensity to a particular density of excited Er$^{3+}$ ions. The number of excited Er$^{3+}$ ions in SiO$_2$ reference sample can be calculated through excitation photon flux when we assume that all Er$^{3+}$ ions are equivalent and they all contribute to the PL process. As has been discussed previously,[6, 10-12] under steady state conditions, the photon flux dependence of Er PL intensity is well described with the formula:

$$I_{PL} \propto N_{\text{Er(SiO}_2)}^* = 100\% \times \frac{N_{\text{Er(SiO}_2)}\sigma_{\text{abs(SiO}_2:\text{Er})}\tau_{\text{rad}}^{\text{Er(SiO}_2)}\Phi}{1+\sigma_{\text{abs(SiO}_2:\text{Er})}\tau_{\text{rad}}^{\text{Er(SiO}_2)}\Phi} \qquad (3)$$

where $\sigma_{\text{abs(SiO}_2:\text{Er})} = 4.17 \times 10^{-22}$ cm$^2$ is the absorption cross-section of Er$^{3+}$ for the $^4I_{15/2} \rightarrow {}^4I_{9/2}$ transition,[13] $\tau_{\text{rad}}^{\text{Er(SiO}_2)} = 14.5$ ms is the radiative lifetime in the excited state, $^4I_{13/2}$,[13] $\Phi$ is the excitation photon flux, and $N_{\text{Er(SiO}_2)} = 9.9 \times 10^{14}$ cm$^{-2}$ is the area concentration of Er ions in the SiO$_2$ reference sample. As shown in the Fig. 1d, the photon flux dependence of PL intensity from SiO$_2$:Er sample shows a linear behavior for low excitation density. Under this condition, *i.e.* $\sigma_{\text{abs}}\tau_{\text{rad}}\Phi \ll 1$, this formula gives a linear dependence on flux: $N_{\text{Er(SiO}_2)}^* = N_{\text{Er(SiO}_2)}\sigma_{\text{abs(SiO}_2:\text{Er})}\tau_{\text{rad}}^{\text{Er(SiO}_2)}\Phi$. Finally, multiple reflections at the interfaces of air-GaN, GaN-air, air-SiO$_2$ and SiO$_2$-air might play a role, we have employed Fresnel's equations for our calculations of photon flux from the Ti:sapphire laser entering the active layer. By substituting this into Eq. 2, we can estimate the number of excited Er$^{3+}$ ions in GaN sample (Fig. 1d). In this way we rescale the right hand side scale of Fig. 1d until the solid red line, the calculated density of excited Er$^{3+}$ ions, overlaps with the PL intensity of MQWs:Er sample.

The PL from the MQWs:Er sample also shows a linear dependence with photon flux. As we mentioned before, under resonant excitation at 809 nm, only the Er optical centers are excited. Thus, the excitation of the Er optical centers in GaN can be calculated from the photon flux and the absorption coefficient under low excitation density:

$$N_{\text{Er(GaN)}}^* = A\% \times \left(N_{\text{Er(GaN)}}\sigma_{\text{abs(GaN:Er)}}\tau_{\text{rad}}^{\text{Er(GaN)}}\Phi\right) \qquad (4)$$

where A is the percentage of Er ions that are Er optical centers in MQWs, $\sigma_{\text{abs(GaN:Er)}} = 3 \times 10^{-20}$ cm$^2$, $\tau_{\text{rad}}^{\text{Er(GaN)}} = 3.3$ ms,[6] $N_{\text{Er(GaN)}} = 1 \times 10^{21}$ cm$^{-3}$, and the total thickness of GaN:Er in the MQW structure of 500 nm. The dash line represents all Er ions in the MQWs being optically active, i.e., or A = 100. From this calculation we have determined the percentage of Er optical centers in MQWs to be 65 ± 5 %, which is high enough to expect optical amplification in these materials. Using the same method, we found that a fraction of ~ 68% of Er ions in the single GaN:Er layer are optically active.[5] This value is higher than that of Eu optically active centers, emitting at 621 nm, in GaN:Eu material that had been determined by a similar approach.[14]

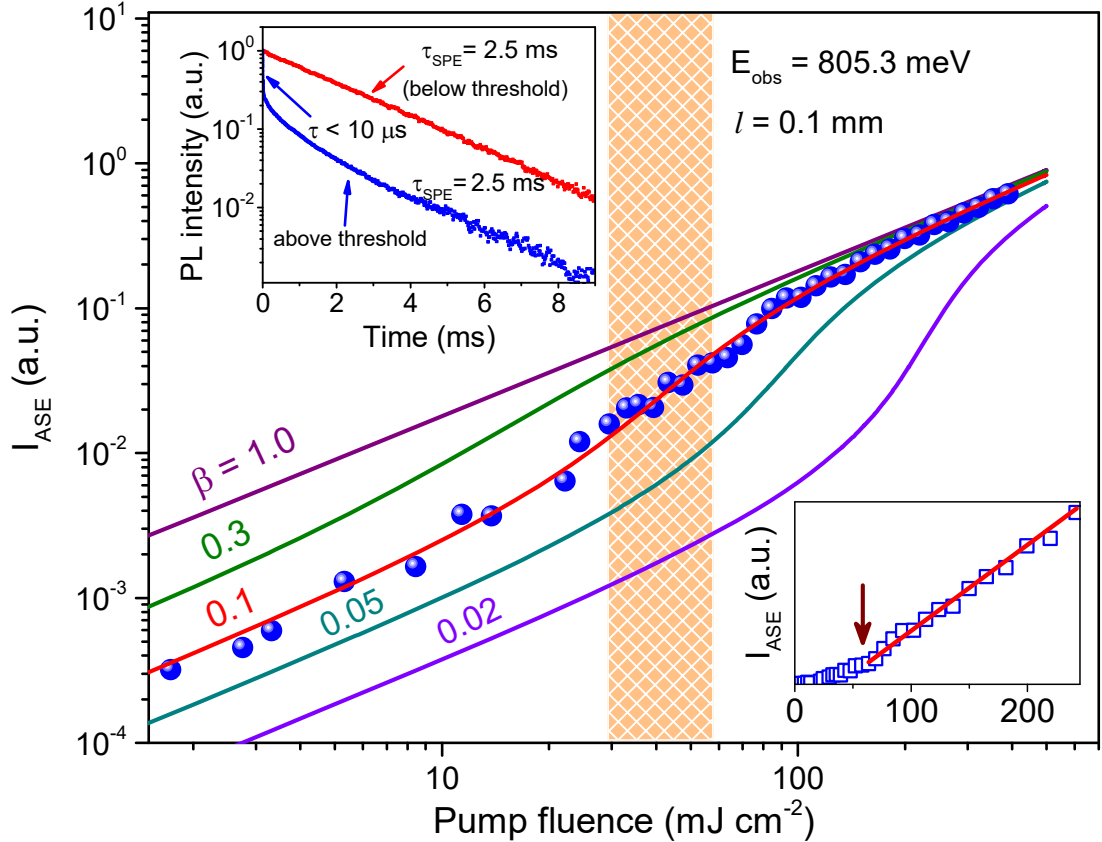

**Figure S4:** L-L data on a log-log scale showing the dependence of the amplified spontaneous emission, $I_{ASE}$, on the pump fluence for the PL peak at 805.2 meV with the excitation length of 0.1 mm. The data were fitted using the S-curve model. Upper inset: Decay dynamics of PL intensity with fluence below (P ~ 0.8$P_{Th}$, red) and above (P ~ 1.5$P_{Th}$, blue) the threshold. Lower inset: Linear plot of the amplified spontaneous emission showing linear behavior below threshold and superlinear increase at higher pump fluence.

### 4. Rate equation analysis

We have employed rate equation analysis to fit the experimental light-in-light-out (L-L) curves, in order to estimate the threshold gain of the MQWs:Er sample with a simple cavity. Figure S4 shows the amplified spontaneous emission, $I_{ASE}$, as a function of the pump fluence for the PL peak at 805.2 meV (1539.80 nm) with the excitation length of 0.1 mm. The linear plot of the amplified spontaneous emission showing linear behavior below threshold and superlinear increase at higher pump fluence (Fig. S4, inset).[15-17]

The rate equations describe the evolution of the carrier density, N, in the active region and the photon density, P, in the cavity mode under optical pumping:[18-21]

$$\frac{dN}{dt} = R_p - \frac{N}{\tau_{sp}} - \frac{N}{\tau_{nr}} - v_g g P \tag{5}$$

$$\frac{dP}{dt} = \Gamma v_g (g - g_{th}) P + \Gamma \beta \frac{N}{\tau_{sp}} \tag{6}$$

where $R_p$ is the optical pumping rate; $\Gamma$ is the confinement factor of the lasing mode; $\beta$ is the spontaneous emission coupling factor; $1/\tau_{sp}$ is the total spontaneous emission rate; $1/\tau_{nr}$ is the non-radiative recombination rate; $v_g = c/n$ is the group velocity; n is the refractive index of the material; g is the material gain; and $g_{th}$ is the threshold gain. A linear relation between the material gain, g, and carrier density, N, is assumed in the active region, $g = \alpha(N - N_{tr})$, where $\alpha$ is a material constant, and $N_{tr}$ is the transparency carrier density. The transparent carrier number can be set to be zero, since it does not affect the fitting result significantly. The rate of non-radiative recombination in the MQWs is currently not known. Any non-radiative decay process would induce additional loss, which would result in a large $\beta$ factor.[22] Our measurements are taken under optical pumping by a 351 nm cw Argon laser at room-temperature. For the steady state solution of the above rate equations, we set $dN/dt = 0$ and $dP/dt = 0$ to obtain.[18-20]

$$R_p = \frac{v_g g_{th} P}{\left(aP + \beta \frac{1}{\tau_{sp} v_g}\right)} \left(aP + \frac{1}{\tau_{sp} v_g}\right). \tag{7}$$

We have fitted the experimental L-L curves with Eq. 7 by first finding the material constants that match the measured threshold pump intensity where a nonlinear jump of the output intensity was observed. Then, the spontaneous emission factor, $\beta$, was varied until the best fit with the experimental data points was obtained. The lasing threshold for the excitation length of 0.1 mm was 55 mJ cm$^{-2}$ of the pump fluence, which corresponds to the best fit for spontaneous emission coupling factor, $\beta$ = 0.1 (Fig. S4). L-L curves for different $\beta$ values are also plotted for comparison, clearly showing the distinct position of the lasing threshold.